\documentclass{article}

\usepackage{graphicx}%
\usepackage{multirow}%
\usepackage{amsmath,amssymb,amsfonts}%
\usepackage{amsthm}%
\usepackage{mathrsfs}%
\usepackage[title]{appendix}%
\usepackage{xcolor}%
\usepackage{textcomp}%
\usepackage{manyfoot}%
\usepackage{booktabs}%
\usepackage{algorithm}%
\usepackage{algorithmicx}%
\usepackage{algpseudocode}%
\usepackage{listings}%
\usepackage{geometry}
\usepackage[square,numbers]{natbib}
\usepackage{authblk}

\title{Direct Visualization of the Magnetic Monopole Field in a 3D Artificial Spin Ice}

\author[1]{Arjen van den Berg}
\author[2]{Peter Rickhaus}
\author[3,4]{Frank Barrows}
\author[4]{Cristiano Nisoli}
\author[1,*]{Sam Ladak}

\affil[1]{School of Physics and Astronomy, Cardiff University, Cardiff, CF24 3AA, United Kingdom}

\affil[2]{Qnami AG, 4132, Muttenz, Switzerland}

\affil[3]{Theoretical Division, Los Alamos National Laboratory, Los Alamos, 87545, NM, United States}
\affil[4]{Center for Nonlinear Studies, Los Alamos National Laboratory, Los Alamos, NM, 87545, United States}
\affil[*]{Corresponding author: Sam Ladak, ladaks@cardiff.ac.uk}

\AtBeginDocument{%
  \newgeometry{a4paper,left=1.5cm,right=1.5cm,top=2.5cm,bottom=2.5cm}%
}
\begin{document}
\maketitle

\abstract{Magnetic monopoles, long hypothesised as fundamental particles carrying isolated magnetic charge, emerge in spin-ice systems as fractionalised excitations governed by the ice rule. Yet their three-dimensional field structure has never been directly visualised. Here, we use two-photon lithography and processing to fabricate a fully three-dimensional artificial spin-ice lattice with diamond-bond geometry. We then use scanning nitrogen-vacancy magnetometry to directly measure the stray magnetic fields of both charge-neutral and monopole vertices. We find that ice-rule vertices produce antivortex textures directly above their vertices, stabilised by the local frustrated two-in/two out ordering principle. Direct imaging of the monopole stray field shows a highly divergent profile. By correlating experiment with micromagnetic simulations and performing a multipole expansion of the reconstructed magnetisation, we reveal that  monopoles in 3DASI are non-trivial micromagnetic entities, carrying both magnetic charge and an intrinsic moment, giving rise to anisotropic interactions that are dependent upon the quasiparticles position on the lattice. Results suggest that as monopoles separate under an applied field, the dipolar contribution to their interaction reorients relative to the underlying Coulombic field, revealing that monopole coupling is tunable through geometry, being set by the local vertex topology. These findings establish 3DASI as a programmable magnetic metamaterial in which nanoscale geometry governs the energetics and dynamics of emergent magnetic charges.}


\section{Introduction}\label{sec1}

While long hypothesized \cite{RN38}, magnetic monopoles have not yet been observed as elementary particles. Nonetheless, they have been proposed \cite{RN15} as emergent classical spinons in spin ices \cite{RN393} such as dysprosium titanate ($\mathrm{Dy_2Ti_2O_7}$) \cite{RN22,RN394} and holmium titanate ($\mathrm{Ho_2Ti_2O_7}$) \cite{RN395}. There, magnetic moments sit on the edges of corner-sharing tetrahedra and are constrained by crystal field anisotropy to point inward or outward. A head-to-tail alignment minimizes interaction energy, but geometric frustration prevents all pairwise interactions from being satisfied, leading to “ice-rule” low-energy configurations where two spins point in and two point out per vertex. This constrained disorder, produces a highly degenerate ground state with residual entropy \cite{RN189}, akin to that of proton-disordered water ice \cite{RN17}. Flipping a spin within a ground state configuration breaks the ice rule and creates a +2/–2 monopole pair on neighbouring vertices, with charge defined as the net number of north minus south poles per vertex \cite{RN15}. Further spin flips separate the monopoles, which can propagate and eventually annihilate with opposite charges. Since their introduction, magnetic monopoles have drawn wide interest for explaining field-induced phases \cite{RN22}, anomalous magnetic noise from their random walks \cite{RN387,RN388}, and other emergent phenomena. They also offer a compelling example of fractionalisation, realising a classical analogue of the so-called Coulomb phase \cite{RN169}. However, magnetic monopoles have not been directly visualised in rare-earth spin ices due to their molecular size. Although, recent proposals outline nitrogen-vacancy (NV) magnetometry as a promising avenue to detect surface monopoles in spin-ice materials.  \cite{RN367}.

The advent of nanotechnological fabrication has allowed the realisation of artificial spin ices (ASIs) \cite{RN105}, arrays of single-domain, interacting Ising-like magnetic nanowires or nanoislands \cite{RN19} which have successfully emulated some of the key physics observed in the bulk counterparts including the Coulomb phase \cite{RN14}. ASIs have also made the visualization and study of magnetic monopoles possible because of the micrometer-scale dimensions of their nanofabricated vertices, but have so far mostly been restricted to two-dimensional arrays, effectively reproducing the six-vertex model on square lattices and not capturing the degeneracy seen in bulk systems \cite{RN232,RN233,RN165}.

Advances in techniques such as multistep electron-beam lithography with precise alignment \cite{RN14}, focused electron beam induced deposition (FEBID) \cite{RN254,RN250} and two-photon lithography (TPL) \cite{RN36,RN163,RN39,TPL_Review} have revolutionized 3D nanomagnetism \cite{RN390} over the past decade, leading to intricate nanostructures \cite{ruiz2025tailoring,skoric2022domain}, unique 3D spin textures, reconfigurable stray fields \cite{RN369}, and distinct topological \cite{Makarov_Flowers,Makarov_EulerChar} and chiral features \cite{RN390}. Recently, these have also led to the realization of the first 3D artificial spin-ice  (3DASI), with offset square systems \cite{RN14}, buckyballs \cite{RN289}, pyramids \cite{berchialla2025magneticfieldcontrolemergentorder}, diamond-bond lattices \cite{RN36,RN126}, their building blocks \cite{RN39,RN252} and gyroids \cite{RN373}. Although the signatures of magnetic monopoles have been individually visualized in 3DASI systems at the surface \cite{RN36,RN126,RN348}, direct measurements of the stray field distribution around monopoles and charge neutral ice-rule vertices remain unreported. NV centre magnetometry offers a promising opportunity for non-invasive characterisation of the stray field around 3D complex nanostructured magnetic systems, a feat that has not yet been accomplished to date.

Here we combine two-photon lithography and thermal evaporation to fabricate a three-dimensional artificial spin ice of diamond-bond geometry. We use scanning NV magnetometry to map the stray magnetic field above the structure under various controlled magnetic configurations. By benchmarking contrast against micromagnetic simulations, we reconstruct the real-space field distribution.  We find that ice-rule vertices, with zero net magnetic charge, produce complex topological stray field textures with features reminiscent of anti-vortices. By applying tailored magnetic field protocols, we nucleate monopole/anti-monopole pairs which allow the direct measurement of their highly divergent field profiles. A clear discrepancy between the observed monopolar field and the idealized field from a point magnetic charge reveals that magnetic monopoles in this system are emergent particles endowed not only with magnetic charge, but also with an intrinsic magnetic moment opening the door to tuneable, anisotropic  Coulomb-like interactions that can be modulated by harnessing sub-100nm 3D nanostructuring \cite{RN362,RN144} and opening the door to controlled monopole dynamics in 3D nanomagnetic systems.

\section{A 3D Artificial Spin-ice}\label{sec2}

\begin{figure}[p]
\centering
\includegraphics{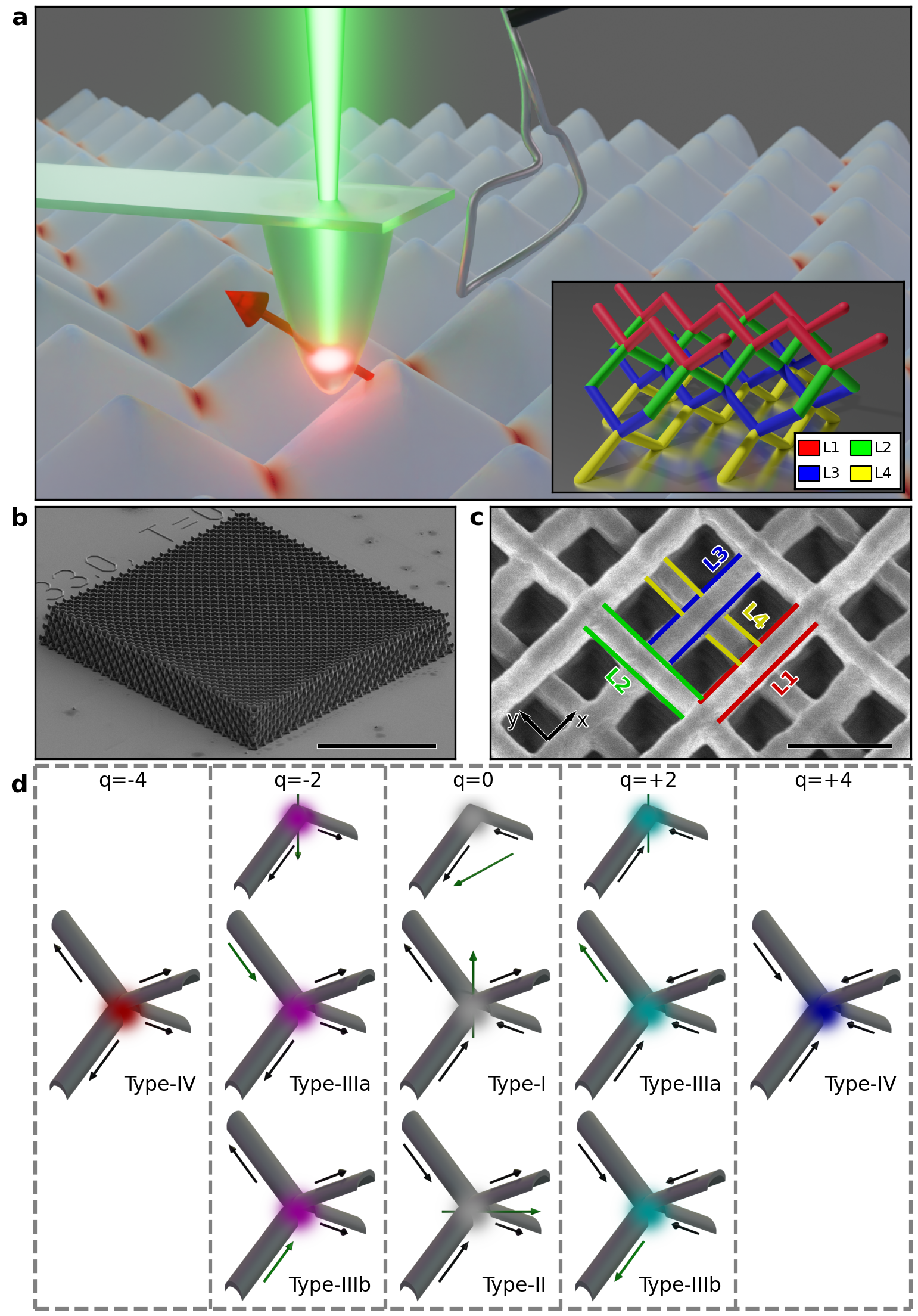}
\caption{\small{ \textbf{A 3D Artificial Spin-ice} 
(a) A diamond probe with a single nitrogen vacancy is scanned across a 3D artificial spin-ice sample and fluorescence is measured with a small antenna. The process measures the component of the sample field along the NV axis indicated by a red arrow. Inset: Schematic of a diamond-bond 3D Artificial Spin-ice lattice with sublattices coloured red (L1), green (L2), blue (L3) and magenta (L4). (b) Scanning electron microscopy image of representative experimental lattice measured at a 45 degree tilt. Scale bar indicates 20 µm. (c) High-magnification top-down SEM of the 3D artificial spin-ice lattice with sublattices marked according to the colours in panel b. Scale bar indicates 1 µm.
(d) Schematic showing different vertex types and corresponding magnetic charge, black arrows indicate each wire’s magnetisation, green arrows indicate the net magnetic moment of the vertex.}}\label{Fig1}
\end{figure}

Scanning NV Magnetometry (SNVM) exploits Zeeman splitting in a single NV defect implanted into a diamond probe which is scanned across a magnetic sample (Fig \ref{Fig1}a) with the extent of Zeeman splitting determined using Optically Detected Magnetic Resonance (ODMR) measurements \cite{RN361}.
A small bias field ($H_b$) offsets the peaks with the sample stray field ($H_S$) perturbing the peak position (Supplementary figure 1), allowing sign information to be recovered provided $H_B>-H_s$. 
We apply the technique to a connected 3DASI lattice with magnetic nanowires arranged along the bonds of a diamond lattice structure with a 2 µm lattice parameter yielding wires of length 886 nm, formed from 3D polymer scaffolds, written in a negative-tone photoresist using TPL, upon which a 50 nm Permalloy ($\mathrm{Ni_{81}Fe_{19}}$) layer is deposited via thermal evaporation \cite{RN36,RN126} (See methods). The structure spans 25 unit cells along the x and y axes, and 1 unit cells along the z axis. Our sample includes 2D rectangular micro-islands with varying aspect ratio to reproduce well-established micromagnetic textures for benchmarking the images against magnetic force microscopy measurements. Supplementary Figure 2 shows that typical vortex and single domain configurations produce contrast that match MFM measurements and the contrast predicted by micromagnetic simulations. 

The inset of figure \ref{Fig1}a shows a schematic of wires in the diamond bond lattice and indicates how one may further divide this structure into sublattices defined by a wire’s z coordinate. On the surface, sublattice L1 (red) consists of wires that alternate between vertices of coordination number two and coordination number four. Sublattice L2 (green) and L3 (Blue) comprise wires connected at coordination number four vertices only. The final sublattice L4 (Magenta) alternates between coordination number two and four, although we note that the coordination number two vertices are disconnected due to shadowing effects of L1 during thermal evaporation. A scanning electron micrograph (SEM) of a representative structure is shown in figure \ref{Fig1}b, with a higher magnification top-view image in figure \ref{Fig1}c further highlighting the division into the sublattices.

Figure \ref{Fig1}d outlines the possible configurations of the magnetisation at each vertex, with the corresponding charge and wire magnetisations associated with each configuration. The charge-free ice-rule vertices carry a net magnetisation along the principal axes, and although the distinction between Type-I and Type-II is typically not meaningful in three-dimensions, the cross-section of our wires still yields some degeneracy lifting among the ice-rule vertices \cite{RN126}. In contrast, the $\mathcal{Q}=\pm4$ vertices with an all-in, all-out configuration, show no net magnetic moment. The $\mathcal{Q}=\pm2$ vertices, analogous to the monopoles in bulk spin-ice, are of particular interest with a net magnetic moment along the outlier spin direction such that one would expect significant deviation from an idealised monopole field. Note, again due to broken symmetry of the wire cross-section, one can define Type IIIa monopoles with outlier spins on the lower wires and Type IIIb monopoles with outlier spins on the upper wires. 

\section{Topological spin textures in stray field above ice-rule vertices }\label{sec3}

\begin{figure}[p]
\centering
\includegraphics{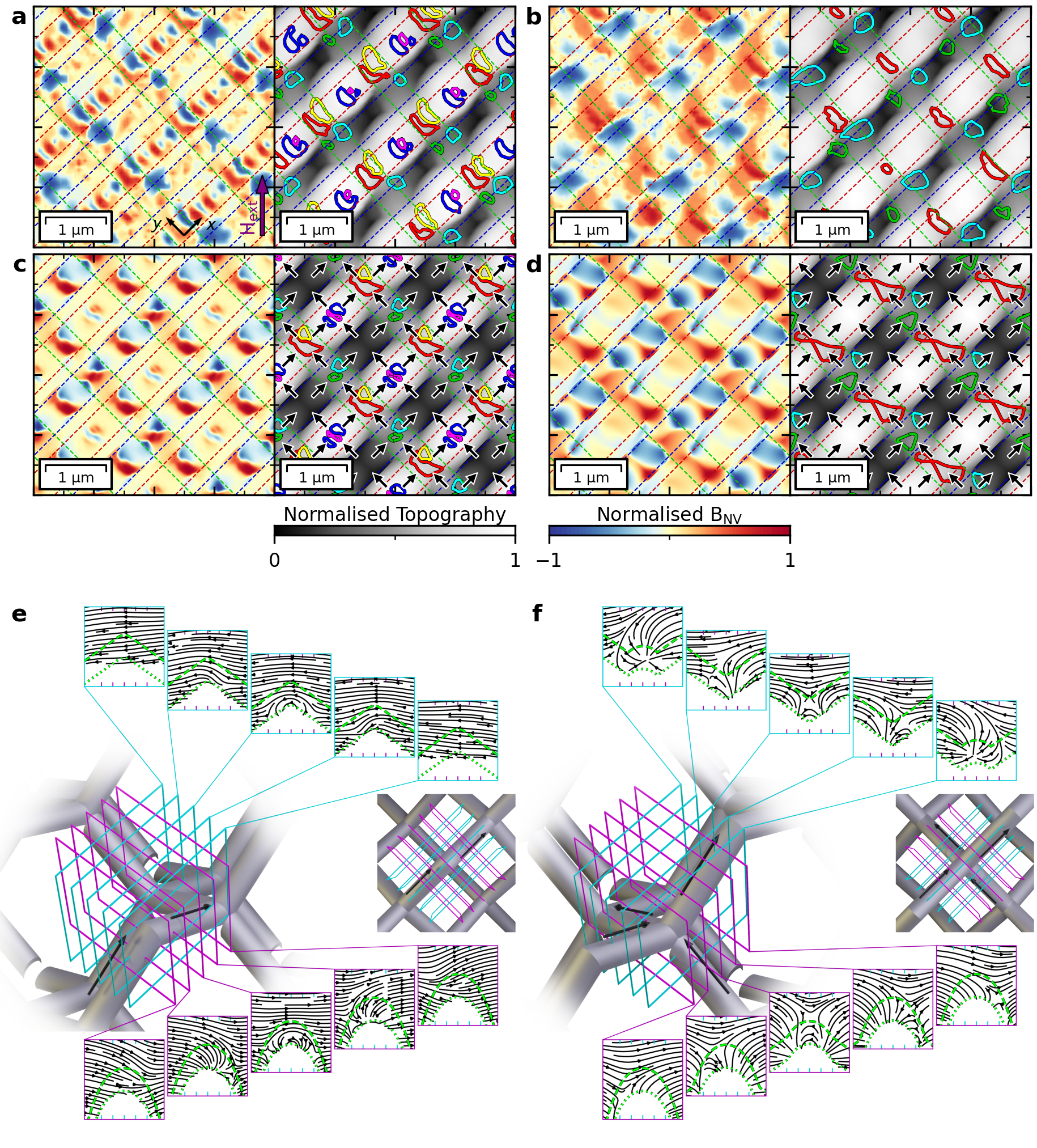}
\caption{\small{\textbf{NV magnetometry imaging of a 3D artificial spin ice lattice in a polarised state}
(a) NV magnetometry image measured in contact mode (left) with measured topography (right). (b) NV magnetometry image measured at 200 nm lift height (left) with measured topography (right). Key features observed in the measured contrast are superimposed on corresponding topography to identify their location. (c) Simulated NV magnetometry contrast in contact mode and (d) at a 200 nm lift height. Key features in the simulation are superimposed on corresponding topography measurements for direct comparison to experiment. Dotted lines in panels a-d indicate the locations of the sublattices to guide the eye, coloured according to the schematic in figure 1a, inset. (e) Tomographic representation of the simulated sample field over the L1 coordination two vertex. Cyan panels show cross-sections of the field in the x-z plane and magenta panels show cross-sections of the field in the y-z plane. In both cases, a circulating texture is present directly above the apex. (f) Tomographic representation of the simulated sample field above the L1/L2 vertex. Cyan panels show cross-sections of the field in the x-z plane and magenta panels show cross-sections of the field in the y-z plane. An antivortex can be seen at the central part of the vertex and arises due to the ice-rule configuration in surrounding nanowires. }}\label{Fig2}
\end{figure}

 We proceed to measure the magnetic field distribution above the 3DASI lattice. A 20 mT field was applied in the substrate plane, at 45$\deg$ relative to the L1 and L2 sub-lattice as depicted in Figure \ref{Fig2}a. Previous work has demonstrated that a field of this magnitude and angle places the system into a polarised state with 2-in/2-out ice-rule states on all coordination-four vertices and 1-in/1-out states on coordination-two vertices \cite{RN126}. The sample was then subject to NV magnetometry imaging whereby the stray field was measured in a lifted two-pass mode resulting in two maps, the first in contact mode (Fig \ref{Fig2}a, left panel) along with the topography (Fig \ref{Fig2}a, right panel) and the second capturing the stray field (Fig \ref{Fig2}b, left panel) at a height of 200 nm above the measured topography (Fig \ref{Fig2}b, right panel). The lift pass also aids interpretation of contact mode measurements adding confidence for identifying sign information and artefacts. We note that the implantation depth of the NV centre was 50 nm, such that the effective lift-height is 50 nm in contact mode and 250 nm on the lift pass. Key features found in the magnetometry data (Fig \ref{Fig2}a \& \ref{Fig2}b, left panels) are overlaid upon the measured topography (Fig \ref{Fig2}a \& \ref{Fig2}b, right panels) to show origin of the contrast. Starting with the contrast observed on the upper surface sub-lattice (L1), which is outlined in dashed red lines, three lobes are observed at the apex of the coordination-two vertices (Fig \ref{Fig2}a, left) and outlined in dark blue/magneta upon the topography image (Fig \ref{Fig2}a, right). The L1/L2 vertex, which can be found at the intersection of dashed red and dashed green lines, hosts elongated lobes of positive and negative contrast across the junction ( red and yellow isolines in Figure \ref{Fig2}a, right panel). Points of intense magnetic field are also observed on the edge of all L2/L3 vertices, which can be found at the intersection of dashed green and blue lines. Here, a larger negative lobe is located at one of the junction corners and a smaller positive lobe at an opposing corner (cyan and green isolines in Fig \ref{Fig2}a, right panel). Lift-pass measurements are shown in Fig \ref{Fig2}b, where it can be immediately seen that fine, opposing features close to the L1 coordination-two vertex are no longer resolved. In contrast, features are still observed at the L1/L2 vertex where the lobe of positive contrast has grown at the expense of the negative lobe and the L2/L3 vertex where contrast appears to broaden. Overall, these variations in contrast suggest the presence of complex position-dependent textures in the sample stray field.

Finite element micromagnetic simulations were performed (See Methods, Supp Fig 3) upon magnetised (Type II tiled) systems, tessellated with relevant stitching at midpoints (Supp Fig 4) and the resultant magnetisation configuration used to calculate the stray field. Figure \ref{Fig2}c, left panel shows the calculated NV contrast 50 nm above a simulated imaging surface obtained from a simple binary search algorithm accounting for interactions between the sample and the NV probe with a finite tip size (See Methods, Supp Fig 5). Crucially, each of the main measured features in the contrast and their position are reproduced and highlighted with the corresponding colours in figure \ref{Fig2}c, right panel, though we note some slight discrepancy in position of contrast on L2/L3 junction which we expect is due to the imperfect calculated NV trajectory for deep valleys in the sample topography. Field calculations were repeated at a height of 250 nm (Fig \ref{Fig2}d), reproducing much of what was observed in the experiment including the absence of clear contrast on the L1 apex and stronger positive contrast at the L1/L2 junction. We note that the experimental contrast upon L2/L3 vertex is not reproduced well by simulation, with broader, offset lobes that span beyond the junction. Again we expect that discrepancies in the calculated topographic pass and the reduced resolution at higher lift height make a clear comparison more difficult. 

Overall, these studies suggest good agreement between experimental and simulated contrast for the upper two sub-lattices and associated junctions and we now proceed to visualise the calculated stray field at positions of key contrast (see Methods). Starting with the contrast above the L1 coordination-two vertex, we extract the field across planes both parallel (magenta axes) and perpendicular (cyan axes) to the nanowire cross-section to produce a tomographic representation of the sample field shown in figure \ref{Fig2}e.
Strongly localised circulating textures are observed directly above the L1 apex through which the NV centre passes in contact mode (Fig \ref{Fig2}e, green dotted line) resulting in the features outlined in blue/magneta on figure \ref{Fig2}a/c, right. These features are not observed in the lift pass as the NV path (Fig \ref{Fig2}e, green dashed line) is well away from the observed circulating texture. Simulations suggest that the origin of this texture is due to the local non-zero surface charge generated by the apex curvature (Supp Fig 6), providing a possible signature of the effective Dzyaloshinskii–Moriya interaction \cite{RN2,RN3}. We note that a canting in the magnetisation transverse to the long axis, also produces a vortex in the plane perpendicular to the wire cross-section, overall yielding a complex 3D texture in the stray field above L1 coordination-two apex. 

Field profile calculations above the L1/L2 vertex (Fig \ref{Fig2}f) reveal the presence of an antivortex directly located over the junction, which can be seen in the cross-sections that are both transverse to the nanowire cross-section (Fig \ref{Fig2}f, cyan) and parallel to the cross-section (Fig \ref{Fig2}f, magenta). In contrast to the profile observed at the L1 apex where the field can be attributed to local magnetic textures, the antivortex above the L1/L2 junction is due to the superposed stray field from the four wires in the tetrapod. This is confirmed and reproduced using simple compass-needle and point-dipole approximations as shown in supplementary figure 8 and 9 respectively, where the antivortex is outlined in blue. 

\clearpage
\section{Measuring Monopolar Fields}\label{sec4}

\begin{figure}[!htbp]
\centering
\includegraphics{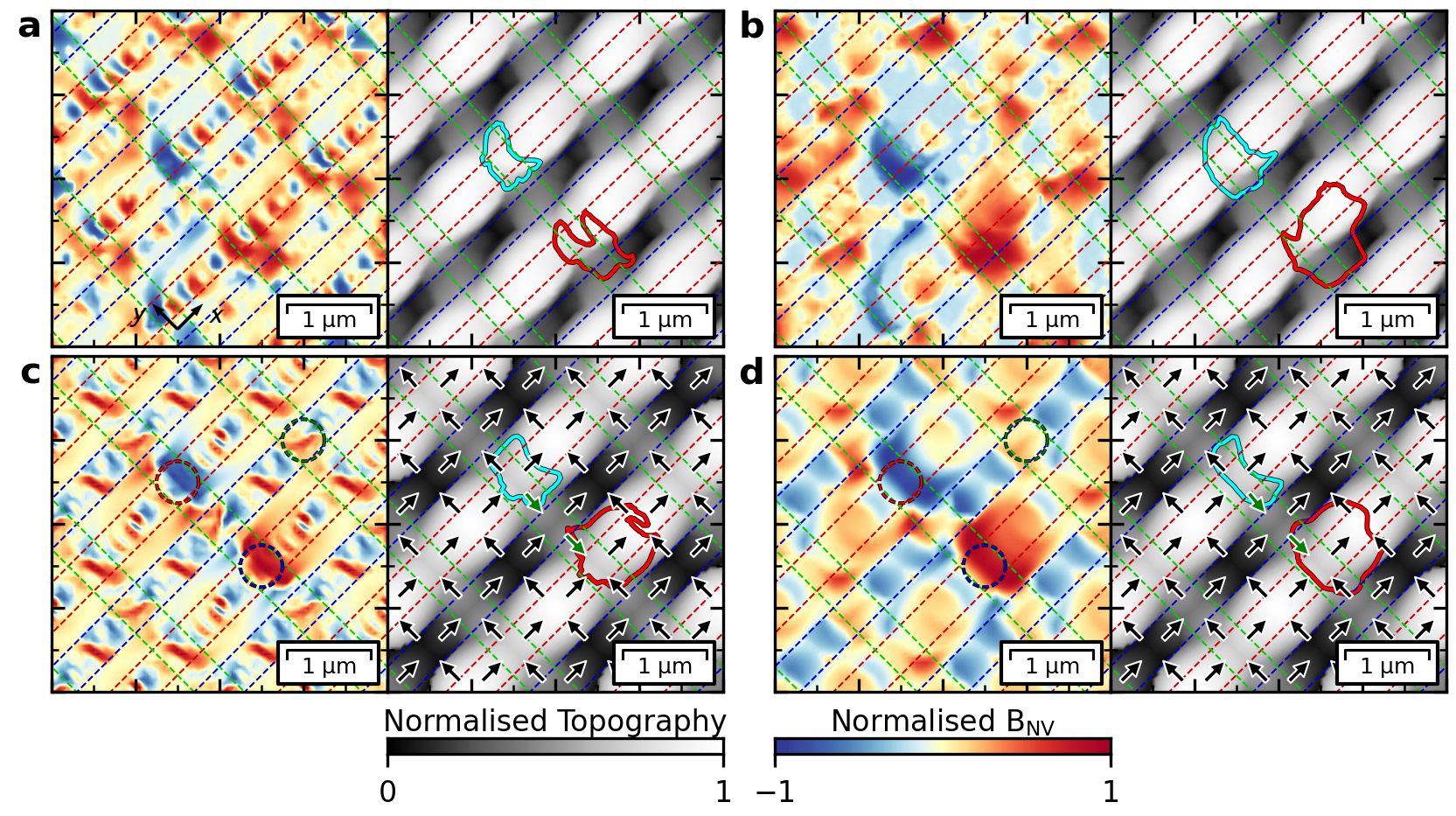}
\caption{\small{\textbf{NV magnetometry of magnetic monopoles in 3DASI} (a) NV magnetometry measurement in contact mode (left) with measured topography (right). A pair of monopoles with bright contrast can be clear seen in the centre of the image. (b) NV magnetometry measurement at 50 nm lift height (left) with measured topography (right). Key features superimposed upon corresponding topography measurements shown in right panels. (c) Simulated NV magnetometry measurements in contact mode (d) Simulated NV magnetometry measurements at a 200 nm lift height with contours of key features superimposed upon corresponding simulated topography measurements (right). Dotted lines in panels a-d indicate the locations of the sublattices to guide the eye, coloured according to the schematic in the inset of figure \ref{Fig1}a.}}\label{Fig3}
\end{figure}

We nucleate monopole pairs by application of a 4.8mT field opposing the initial saturating field. Figure \ref{Fig3}a, left shows the contrast corresponding to a pair of monopoles captured in the contact mode pass, with their locations highlighted in the corresponding topography measurement (Fig \ref{Fig3}a, right). Here the red isoline corresponds to a 3-in/1-out positive monopole and the cyan isoline corresponds to the 1-in/3-out negative monopole. The contrast associated with these magnetic charges becomes broader on the lift pass shown in figure \ref{Fig3}b, left, consistent with a highly divergent stray field. By tracing key NV contrast isolines that deviate from the Type-II background onto the topography measurements one can show that the monopoles are located upon L1/L2 vertices, enabling an approximation of the NV contrast to be calculated based on outputs from micromagnetic simulations, with the contact-mode pass shown in Fig \ref{Fig3}(c), left and lift-mode pass shown in Fig \ref{Fig3}d, left. The isolines extracted from the experimental contrast map reasonably well onto the simulated contrast, although some deviation is present in the contrast associated with the positive charge. Notably, the red isoline in figure \ref{Fig3}a, right shows a small region of negative contrast and does not display the symmetry present in the simulated contrast. This asymmetry may be due to the superposition from fields created at nearby buried vertices or due to subtle physical disorder such as roughness that cannot be captured in our idealised micromagnetic simulations. 

\begin{figure}[!htbp]
\centering
\includegraphics[scale=0.9]{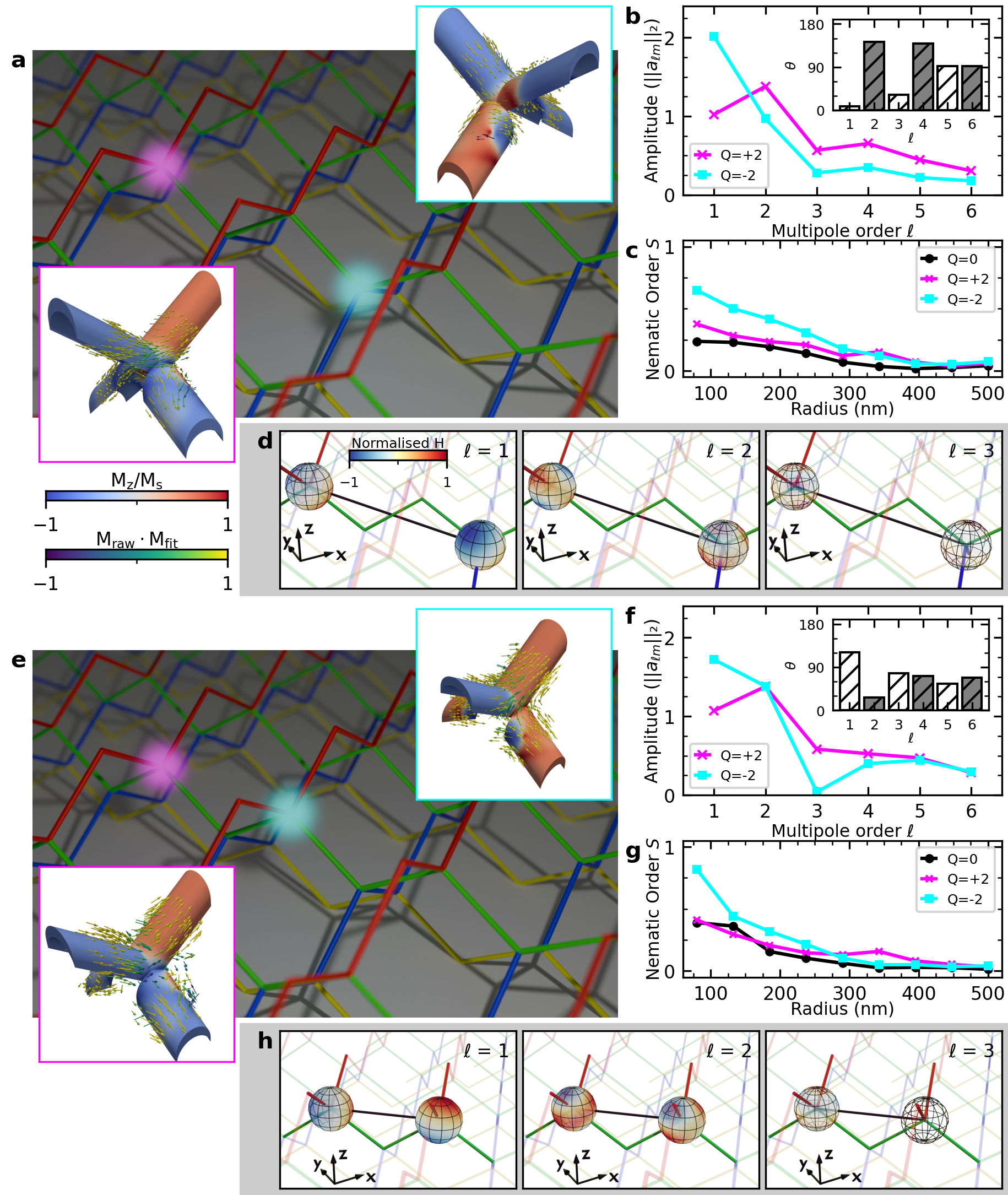}
\caption{
\small{\textbf{Monopole interactions in 3DASI.}
(a–d) Multipole expansion of a Type~IIIa–Type~IIIb monopole pair. (a) Schematic showing monopole positions on the lattice; insets show reconstructed magnetization at a radius of $180 \pm 100$~nm for vertices with $\mathcal{Q} = \pm 2q$. Colour indicates the $z$-component of magnetization, while vectors show the reconstructed in-plane components; vector contrast reflects cosine similarity between reconstructed and raw magnetization. (b) Amplitude of each multipole order with inset showing charge alignment per order. (c) Nematic order $S$ of magnetization for charged (cyan, magenta) and uncharged (black) vertices as a function of radius, where $S=1$ denotes full anisotropy. (d) Low-order multipole moments of the magnetic charges, with real-space orientations indicated. The inter-charge distance is normalised to a unit vector (black line); surface opacity represents the amplitude of multipole order $\ell$. (e–h) Equivalent analysis for a Type~IIIa–Type~IIIa monopole pair.}}

\label{Fig4}
\end{figure}

A central question is how closely the emergent field of a monopole in a three-dimensional artificial spin ice (3DASI) resembles that of an ideal magnetic charge. Although the global condition $\nabla \cdot \mathbf{B} = 0$ must hold, the local magnetisation \textbf{M} within each vertex is highly non-uniform, giving rise to a finite dipolar component. In contrast to the point-like excitations of bulk spin ice, monopoles in 3DASI are micromagnetic objects with structured spin textures confined to the vertex region. The broken symmetry of the nanowire cross-section further lifts the degeneracy between equivalent charge states, producing two distinct monopole types (Type IIIa and Type IIIb) with slightly different energies and topologies (Fig \ref{Fig1}d). Consequently, the effective interaction between monopoles depends not only on their charge and separation, but also on their micromagnetic type and sublattice position.

To quantify how the vertex topology shapes the monopole field, we perform a multipole expansion of the simulated magnetisation using vector spherical harmonics (See Methods). Each vertex was analysed independently, sampling \textbf{M(r)} on concentric shells of 200 nm thickness to capture the three-dimensional spin texture with increasing distance from the core. This approach provides a compact description of the magnetic ordering, separating the dominant monopolar term from higher-order dipolar and lattice-specific contributions. Although the simulations consider individual, non-interacting vertices, the resulting multipole coefficients reveal how broken symmetry and local anisotropy modify the nominal Coulomb field, offering insight into the effective interactions that would emerge in a fully coupled lattice.

We first analyse a Type  IIIa–Type  IIIb monopole pair of opposite charge, representative of monopoles located at junctions connecting different sublattices, for example, one at an L1/L2 junction and the other at an L2/L3 junction as depicted in Fig \ref{Fig4}a. The reconstructed magnetisation fields of each vertex are shown in Fig \ref{Fig4}a insets, yielding low error with respect to the raw simulations. Fig \ref{Fig4}b reveals that the lowest odd multipoles ($\ell = 1$, $\ell = 3$) are co-oriented for the +2 and –2 charges, consistent with a field directed along the connecting vector between the monopoles, while even orders ($\ell = 2$, $\ell = 4$) remain largely orthogonal, reflecting local anisotropy rather than long-range alignment. The nematic order parameter (S), shown in Fig \ref{Fig4}c, further shows that charged vertices exhibit stronger anisotropy (S $\approx 1$ near the vertex), whereas neutral vertices relax into flux-closure states with lower S. Figure \ref{Fig4}d plots the relative orientation of the low order vector spherical harmonic expansion, further illustrating that the low-order multipoles are oriented along the unit vector connecting the two charges. Higher order multipoles are shown in Supp fig 10. This ordered alignment of the odd-parity terms constitutes the microscopic fingerprint of a Coulomb-like interaction, directly linking the vertex spin texture to the directionality of the emergent field.

We next examine a Type  IIIa – Type  IIIa monopole pair of opposite charge, corresponding to monopoles located on equivalent L1/L2 junctions, directly analogous to the experimentally observed configuration and shown schematically in Fig \ref{Fig4}e, with reconstructed magnetisation field shown in the Fig \ref{Fig4}e insets.
The corresponding multipole amplitudes (Fig \ref{Fig4}f) reveal a striking contrast to the mixed Type  IIIa–Type  IIIb case. The lowest odd orders ($\ell = 1$, $\ell = 3$) are now nearly orthogonal, suggesting the absence of long-range alignment or emergent Coulomb-like behaviour.  The nematic order (Fig \ref{Fig4}g) remains high near each vertex but lacks coherent orientation between them. This breakdown of co-alignment originates from the bond geometry. Each vertex’s magnetisation follows the tangent of its local Dirac-string segment, so the dominant dipole axis reflects the incident angle of the connecting nanowires. The resulting orthogonal dipolar orientation captures how lattice symmetry alone can suppress long-range coupling between same-type monopoles.
The contrast between the two monopole pairings arises directly from lattice geometry.
In the Type  IIIa–Type  IIIb configuration, the Dirac string is symmetric under spatial inversion through its midpoint, meaning that both monopoles experience the same local bond orientation on their respective vertices. As a result, the local magnetisation tangents at the two vertices are nearly parallel, giving rise to a Coulomb-like, co-aligned dipolar configuration. 

In the Type  IIIa–Type  IIIa case (Fig \ref{Fig4}h), the connecting bonds are asymmetric and subtend an angle of roughly $70\deg$, producing opposite out-of-plane cant of the local string segments. Consistent with this, the average magnetisation directions on a 200 nm shell differ by $\approx55\deg$ for Type  IIIa–Type  IIIa pairs but only $\approx5\deg$ for mixed pairs, confirming that vertex magnetisation largely follows the incident bond geometry. Higher-order multipoles capture subtle deviations from this leading dipolar order. Overall, the odd-parity terms encode the emergent Coulomb coupling, while the even orders reflect local anisotropy, together forming the multipolar fingerprint of monopole interaction strength and geometry in 3DASI.

These findings reveal that monopole interactions in 3DASI are not fixed but dynamically modulated by lattice geometry. As a monopole pair propagates through the lattice under an applied field, its path alternates between Type  IIIa–Type  IIIb and Type  IIIa–Type  IIIa configurations, leading to periodic reorientation of the dipolar component that is superposed on the underlying Coulombic field. The resulting interaction thus evolves continuously: the monopolar term defines the overall attraction or repulsion, while the dipolar anisotropy imposes a geometry-dependent modulation. The strength and directionality of this coupling therefore depend not only on charge separation but also on the local micromagnetic structure of the traversed vertices. This tunable, geometry-controlled superposition of monopolar and dipolar interactions underscores the metamaterial character of 3DASI, where tailoring vertex shape and cross-section enables control over the energetics and dynamics of emergent magnetic charges.

\section{Conclusions }\label{sec5}

We have directly visualised the stray magnetic field above a 3D artificial spin ice (3DASI) lattice in multiple micromagnetic configurations. Ice-rule vertices at coordination-two and coordination-four junctions give rise to distinct topological features in the stray field with vortices and antivortices, while the nucleation of monopole–antimonopole pairs reveals highly divergent field distributions characteristic of magnetic charges. Although monopoles are typically modelled as isotropic Coulombic defects, multipole analysis shows that in 3D artificial systems each vertex hosts monopoles, with anisotropic and geometry-dependent interactions.

Importantly, these interactions are dynamic and geometry-dependent. As monopole pairs separate under an applied field, they alternate between Type A–Type B and Type A–Type A configurations, producing a periodic reorientation of the dipolar contribution superposed on the underlying Coulombic field. The effective interaction landscape therefore evolves with monopole motion, governed by local vertex topology and spin texture.

This behaviour establishes 3DASI as a reconfigurable magnetic metamaterial, where engineered vertex asymmetry enables direct control over monopole energetics, propagation, and collective ground states. Our approach transforms frustrated 3D spin systems into a quantitatively programmable platform for studying and harnessing emergent quasiparticle dynamics in artificial matter.

\section{Methods}\label{sec6}

\subsection{Sample Fabrication}\label{subesc6-1}
The structures outlined in this work were fabricated by creating polymer scaffolds using two-photon lithography (TPL) upon which a 50 nm permalloy ($\mathrm{Ni_{81}Fe_{19}}$) was deposited using thermal evaporation at a base pressure of 1 x $10^{-6}$ mbar 

The scaffolds were written upon glass cover slips which were cleaned using 20-minute acetone ultrasound bath followed by a 20-minute isopropanol ultrasound bath and dried using compressed air. Index-matched immersion oil was drop-cast on one side of the cover slip, and Nanoscribe’s proprietary IP-L negative tone photoresist cast upon the other side before mounting the samples onto the TPL apparatus. After exposure, excess resist was removed using a 20-minute propylene glycol monomethyl ether acetate (PGMEA) developer bath followed by a 20-minute isopropanol bath and a gentle drying process using compressed air.

\subsection{NV Magnetometry}\label{subsec6-2}
Nitrogen-Vacancy (NV) magnetometry measurements were performed using a Qnami ProteusQ microscope with a high performance Qnami MX+ tip which features an NV flying distance of approximately 50 nm.
The distance is regulated by a tuning fork-based frequency-modulated PID loop, while the magnetic signal is given by the NV fluorescence via optically detected magnetic resonance (ODMR), thus there is no crosstalk between topographic and magnetic information.

\subsection{Magnetic Force Microscopy}\label{subsec6-3}
Magnetic force microscopy (MFM) measurements were performed using a Bruker Dimension ICON scanning probe microscopy system with NANOSENSORS super-sharp low-moment probes.

\subsection{Computational Methods}\label{subsec6-4}
Magnetic textures were computed using both finite difference (FDM) and finite element methods (FEM), selecting material parameters $M_s = 0.86\times10^6 \mathrm{A m^{-1}}$ and $A = 13\times10^{12} \mathrm{J m^{-1}}$ typical for Permalloy.

\subsubsection{Finite difference simulations}\label{subsubsec6-4-1}
2D structures were simulated using the GPU accelerated finite difference package mumax3 \cite{RN77}.
Initial conditions were set as a random magnetisation and the lowest energy state was found using the ‘relax’ command employing an RK23 (Bogacki-Shampine) solver.

\subsubsection{Finite element simulations}\label{subsubsec6-4-2}
Simulations of the various vertex types were performed using the finite element nMag code \cite{RN78} where finite element methods are more suitable for curved geometries.
The simulation geometries were determined by approximating a 200 nm wide polymer scaffold as produced by TPL, and extruding the top surface 50 nm along the z axis to yield magnetic nanowires with a crescent-shaped cross section. Geometries were defined using Blender and exported to a stereolithography (STL) file and meshes created using NetGen software.

The magnetisation of the 250 nm regions at the extremes of each 886 nm long wire were set to be oriented either towards or away from the vertex reflective of the desired vertex type and subsequently frozen.
The magnetisation of the central region was set to random and allowed to evolve to the lowest energy state until convergence (using nMag’s default value) is reached.

\subsubsection{Field and image surface calculations}\label{subsubsec6-4-3}
Approximations of the magnetic configurations shown in Figure \ref{Fig2} and figure \ref{Fig3} were created by combining FEM simulation outputs. The sample field for any position ($\mathbf{R}$) on the imaging surface was calculated using a dimensionless expression for the field due to a point dipole at position $\mathbf{r'}$:
\begin{equation}\label{eqn1}
\mathbf{H(R)}=\sum_i \frac{3 \mathbf{r}_i (\mathbf{r}_i \cdot \mathbf{m}_i)}{r_i^5} - \frac{\mathbf{m}_i}{r_i^3}
\end{equation}

Where $m_i$ indicates the $i^{th}$ dipole’s reduced magnetisation and $\mathbf{r}_i = \mathbf{R}-\mathbf{r^i}_i$.

The SPM imaging surfaces were computed by modelling a parabolic tip with an 80 nm radius at the NV centre implantation depth of 50 nm. The appropriate tip radius was determined phenomenologically by comparing calculated surfaces with measurements. For each x and y coordinate, a binary search identified the greatest extent by which the probe may be lowered onto the sample, without contacting the sample at any point along the probe surface (see supplementary figure 5).

\subsubsection{Dipole and compass needle approximations}\label{subsubsec6-4-4}
Point-dipole and compass needle approximations do not account for micromagnetic textures, as such, these approaches were used to determine whether the topological defects could be attributed to specific micromagnetic textures or the bulk magnetisation of the wires. A single unit cell was constructed using either a point dipole at the centre of each carbon-carbon bond, or a compass needle model with pair of charges some distance from the centre of the bond. Where the distance between adjacent vertices is described with the parameter a, the distance between the charges is described with the parameter b, with the ratio b/a having previously shown to be a useful parameter for compass needle approximations. The unit cell was subsequently tessellated in x and y such that the edges of the system are far from the region of interest.
The field was calculated for planes perpendicular and parallel to the L1 sublattices with the planes centred upon the vertices of interest. The point dipole models use equation \ref{eqn1}, whereas the compass needle takes the sum of the monopole field due to each charge $\mathcal{Q}_i$ at position $r'_i$:
\begin{equation}\label{eqn2}
\mathbf{H(R)}=\sum_i \frac{\mathcal{Q}_i}{r_i}\hat{r_i}
\end{equation}

\subsubsection{Multipole analysis of three-dimensional magnetization}

Magnetization vectors $\mathbf{M}(\mathbf{r})=(M_x,M_y,M_z)$ were obtained on (i) regular simulation grids or (ii) as scattered surface samples (tables of $[M_x,M_y,M_z,x,y,z]$). Scattered data were tri-linearly interpolated onto a Cartesian grid $\{x_i,y_j,z_k\}$.

Around each vertex (centroid $\mathbf{r}_0$) we sampled $\mathbf{M}$ on thin spherical shells of mean radius $R$ and thickness $\Delta R=200$\,nm. Shell radii were linearly spaced (typically $K_{\rm shells}=100$) over $[R-\Delta R/2,R+\Delta R/2]$. 
At each valid direction, we expanded the field on the sphere using the standard radial/poloidal/toroidal VSH bases,
\begin{equation}\label{eqn3}
\mathbf{M}(\hat{\mathbf{u}})\;\approx\;\sum_{\ell=0}^{\ell_{\max}}\sum_{m=-\ell}^{\ell}
\Big[a^{(r)}_{\ell m}\,\mathbf{Y}^{(r)}_{\ell m}(\hat{\mathbf{u}})+
     a^{(e)}_{\ell m}\,\mathbf{Y}^{(e)}_{\ell m}(\hat{\mathbf{u}})+
     a^{(m)}_{\ell m}\,\mathbf{Y}^{(m)}_{\ell m}(\hat{\mathbf{u}})\Big]
\end{equation}
with scalar harmonics $Y_{\ell m}$ and $\ell_{\max}=6$. Coefficients were obtained by Tikhonov-regularized least squares,
\begin{equation}\label{eqn4}
\min_{\mathbf{a}}\;\|\mathbf{Y}\mathbf{a}-\mathbf{b}\|_2^2\;+\;\lambda\,\|\mathbf{a}\|_2^2
\end{equation}
where $\mathbf{b}$ stacks the sampled $\mathbf{M}$ over directions/shells and $\mathbf{Y}$ is the assembled VSH design matrix. Fits were performed independently for each vertex and radius; “good” radii were selected by NRMSE thresholds and used for averaging. The root-mean-square error (RSE) between reconstructed and raw magnetisation remains low, confirming the fidelity of the expansion (Supp Fig 10).

For a given $\ell$, we quantified (i) the rotation-invariant power $\|a^{(r)}_{\ell m}\|_{m}^2$ and (ii) the alignment angle between two vertices by the angle between their coefficient subvectors $(a^{(r)}_{\ell m})_{m=-\ell}^{\ell}$ after unit normalization. As a complementary frame-free readout, a local “dipole axis” $\hat{\mathbf{p}}$ was extracted on a single shell by regressing $s(\hat{\mathbf{u}})=\hat{\mathbf{u}}\cdot\mathbf{M}(\hat{\mathbf{u}})$ onto $\hat{\mathbf{u}}$; $\hat{\mathbf{p}}$ provides a robust leading-axis proxy without relying on basis choice.

The anisotropy of $\mathbf{M}$ on a shell $\Delta R$ was summarized by the traceless second moment

\begin{equation}\label{eqn5}
\mathcal{Q}_{ij}=\Big\langle \hat{u}_i^{\,k}\hat{u}_j^{\,k}-\delta_{ij}/3 \Big\rangle_{k\in\Delta R},\qquad
S(R)=\tfrac{3}{2}\,\lambda_{\max}\!\left(\mathcal{Q}\right)
\end{equation}
with $\hat{\mathbf{u}}^{\,k}=\mathbf{M}^{k}/\|\mathbf{M}^{k}\|$ and $\lambda_{\max}$ the largest eigenvalue.

\section*{Supplementary information}

If your article has accompanying supplementary file/s please state so here.

\section*{Declarations}
\subsection*{Funding}
SL gratefully acknowledges funding from the Engineering and Physics Research Council (EP/L006669/1, EP/R009147) and Leverhulme Trust (RPG-2021-139)

\subsection*{Data Availability}
Information on the data presented here, including how to access them, can be found in the Cardiff University data catalogue. 

\subsection*{Code Availability}
All codes utilised within this study is available upon reasonable request to the corresponding author.
\subsection*{Author Contributions}
SL conceived and supervised the overall project. AV carried out sample fabrication and characterisation using SEM and MFM. NV magnetometry was carried out by AV and PR. Micromagnetic simulations were carried out by AV, as well as developing protocols and code for simulating NV magnetometry measurements.
FB carried out the multipole expansion and subsequent analysis thereof, under the supervision of CN. The initial draft of the manuscript was written by AV, SL, and CN. All authors contributed to the editing of the final manuscript.
\subsection*{Conflict of interest} 
The authors declare no conflict of interest.

\bibliography{Bibliography}
\end{document}